\documentclass[preprint,pra,superscriptaddress,amsmath,amssymb]{revtex4-2}
\usepackage{amsmath}
\usepackage{amsfonts}
\usepackage{amssymb}
\usepackage{graphicx}
\usepackage{epstopdf}
\usepackage{float}
\usepackage{dblfloatfix}
\usepackage{color}
\usepackage{txfonts}
\usepackage[titletoc]{appendix}
\usepackage[colorlinks={true}]{hyperref}
\hypersetup{citecolor={blue}, filecolor={blue}, linkcolor={blue}, urlcolor={blue}}

\begin{document}

\title{Field-induced quantum interference of inelastic scattering in ultracold atomic collisions}

\author{Ting Xie}%
\affiliation{Hefei National Laboratory, University of Science and Technology of China, Hefei 230088, China }%
\affiliation{Shanghai Research Center for Quantum Science and CAS Center for Excellence in Quantum Information and Quantum Physics,
University of Science and Technology of China, Shanghai 201315, China}
\author{Chuan-Chun Shu}%
\email{cc.shu@csu.edu.cn}%
\affiliation{Hunan Key Laboratory of Nanophotonics and Devices, School of Physics, Central South University, Changsha 410083, China}

\date{\today}

\begin{abstract}
Exploiting quantum interference remains a significant challenge in ultracold inelastic scattering. In this work, we propose a method to enable detectable quantum interference within the two-body loss rate resulting from various inelastic scattering channels. Our approach utilizes a ``ring-coupling" configuration, achieved by combining external radio-frequency and static electric fields during ultracold atomic collisions. We conduct close-coupling calculations for $^7$Li-$^{41}$K collisions at ultracold limit to validate our proposal. The results show that the interference profile displayed in two-body loss rate is unable to be observed with unoptimized external field parameters. Particularly, our findings demonstrate that the two-body loss rate coefficient exhibits distinct constructive and destructive interference patterns near the magnetically induced $p$-wave resonance in the incoming channel near which a rf-induced scattering resonance exists. These interference patterns become increasingly pronounced with greater intensities of the external fields. This work opens a new avenue for controlling inelastic scattering processes in ultracold collisions.
\end{abstract}


\maketitle
\section{Introduction}
Quantum interference is a fundamental phenomenon in quantum mechanics, where particles such as electrons, photons, and atoms exhibit wave-like properties \cite{book1,prl:123:223202,nc:8:14854,nat:391:263,prl:122:253201}. This wave-like behavior enables particles to interfere with themselves or with one another, a process that is crucial for determining the likelihood of finding particles in specific states or locations. Such interference highlights the intricate and complex nature of quantum systems and serves as an essential foundation for numerous technological innovations and theoretical models across various scientific disciplines \cite{nat_n:19:986,pnas:121:e2306953121,science:300:1730,sciadv:5:9674,nc:12:1317,pra:87:033607}.
The observation of interference patterns in cold and ultracold collisions and chemical reactions offers invaluable insights into the subtle interplay of quantum mechanical forces \cite{csr:51:1685,science:375:1006,pccp:23:5096,arxiv:2310:07620,science:326:1683}. Specifically, exploring quantum interference phenomena in both elastic and inelastic scattering not only enriches our fundamental understanding of quantum mechanics but also unveils potential applications for the quantum control of chemical reactions. Recent advancements in this field have illuminated the fascinating possibility of exerting coherent control over chemical reactions using superposition states under cold or ultracold conditions \cite{prl:77:2574,jcp:113:2053,pra:78:054702,prl:121:073202,prl:126:153403,prr:5:L042025,jctc:17:7822}. Although this technique is still in its infancy in the context of inelastic scattering, where the challenges of constructing effective multiple pathways persist, it holds the promise of enabling unprecedented control over reaction dynamics. This can lead to the discovery of alternative pathways for manipulating and controlling ultracold collisions at the quantum level.\\ \indent 
In ultracold collision experiments, inelastic scattering plays a crucial role in modifying collision properties but also presents a double-edged sword \cite{rmp:82:1225}. As a probing mechanism, inelastic processes facilitate measurements of Feshbach resonances, molecular binding energies, and dynamic structures \cite{pra:70:032701,prl:116:205303,pra:95:022715,nature:600:429,pra:87:033611,pra:85:032506}. However, they also significantly limit the ability to tune interactions between atoms or molecules at high energy levels \cite{pra:101:052710,prl:125:153202,njp:23:125004,prr:2:033163}. Consequently, exploring and controlling inelastic processes on a quantum level has emerged as a primary goal within the rapidly developing field of ultracold chemistry. In this work, we propose a method for observing both constructive and destructive interference phenomena in ultracold two-body inelastic collisions, without the necessity of employing a coherent state. Unlike most ultracold atomic collision experiments that utilize individual magnetic fields to modulate interactions via Feshbach resonances, our approach employs external radio-frequency (rf) and electric fields, as illustrated in in Fig.  \ref{fig1}.   Ignoring the weak spin-dependent interactions at the ultracold limit, the magnetic field constrains the collisions within a single spin-exchange block (\(M_F\)).  The external fields couple different spin-exchange blocks (\(M_F\)) and partial waves (\(\ell\)), thereby creating a series of ring-coupling structures.  This ring coupling enables the external fields to link adjacent channels effectively.  The field-induced inelastic transitions from one channel in \(M_F\) to a lower energy level in \(M_F \pm 1\) generate multiple scattering pathways.\\ \indent When external field parameters are not optimized, interference among these pathways can hardly modify the overall two-body loss rate. However, by appropriately modulating these external field parameters, we demonstrate that the interference effect can significantly revise the total two-body loss profile at magnetic-induced \(p\)-wave Feshbach resonances in incoming channel. To illustrate our proposal, we use the \( ^7 \text{Li}-^{41} \text{K} \) colliding system as an example, based on several key considerations. First, the relatively light masses of both particles allow us to neglect weak anisotropic spin-dependent interactions, enabling us to focus on the ring-coupling structures generated by external fields \cite{prl:100:053201,epjd:65:55}. Second, the substantial permanent electric dipole moment of the ultracold LiK molecule permits experimentalists to employ practical electric fields to create observable interference patterns \cite{jcp:57:1487,jcp:122:204302,prl:132:243401}. Additionally, the significant difference in scattering lengths between the ground singlet and triplet states can lead to a broad magnetically induced \( p \)-wave resonant profile, which is crucial for tracing the interference pattern in experiments \cite{pra:98:042708}.  This finding holds potentials for controlling the inelastic scattering of ultracold molecules. \\ \indent
This paper is organized as follows: In Sec. \ref{TM}, we present the scattering theory of two colliding atoms in the presence of external fields. In Sec. \ref{RD}, we conduct numerical simulations of ultracold \( ^7\text{Li}-^{41}\text{K} \) collisions to investigate the possibility of observing interference patterns in the two-body loss rate. We summarize our findings in Sec. \ref{Con}.
\section{Theoretical Methods}\label{TM}
\begin{figure*}[h]
\centering
\resizebox{1.0\textwidth}{!}{%
\includegraphics{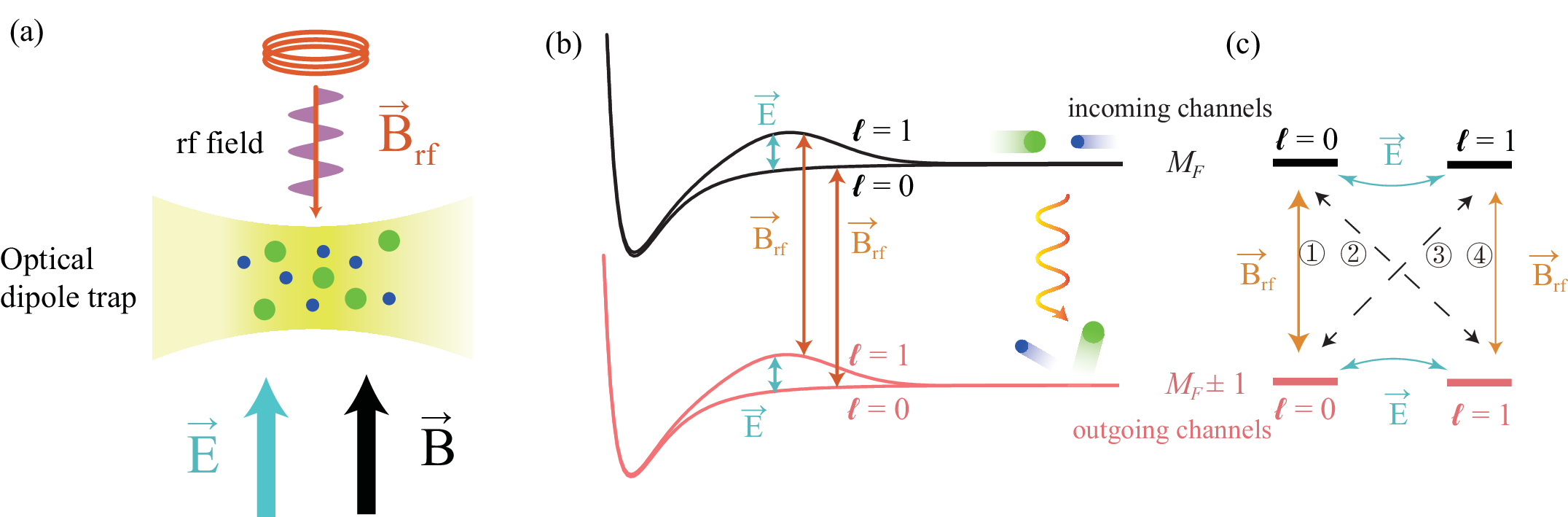} }\caption{ (a) The created ultracold mixtures of two species confined in an optical dipole trap in the presence of magnetic field. The colliding particles of interspecies have an electric dipole moment and magnetic moment, thus can be manipulated with electric and rf fields. (b) A scheme of the coupling system in the presence of external rf and electric fields which directed along the magnetic field. The incoming and outgoing channels, corresponding to different spin-exchange blocks, can be coupled together by rf field $\vec{B}_{\rm rf}$ with the same $\ell$. The electric field $\vec{E}$ couples the channels with different partial waves restricted in an individual $M_F$. The inelastic collision can occur if the atoms in the incoming channels decay to outgoing ones steered by rf field. (c) The direct couplings induced by fields only appear between adjacent channels (solid lines). The transitions between non-adjacent channels (dashed lines) can occur through two dependent indirect couplings. The total two-body inelastic scattering can be decomposed into four independent routes labeled with $\normalsize\textcircled{\small{1}}$ $\thicksim$ $\normalsize{\textcircled{\small{4}}}\normalsize$.} \label{fig1}
\end{figure*}
Figure \ref{fig1}(a) illustrates our model for investigating the behavior of two distinct ultracold atoms confined within an optical dipole trap and subjected to three fields. As shown in Fig. \ref{fig1}(b), the magnetic field confines collisions conserved with \( M_F \), while the rf field interacts with various \( M_F \) states by absorbing and emitting photons \cite{pra:80:050701,njp:12:083031,jpb:45:145302}. Inelastic collisions induced by the rf field may occur when colliding atoms from the incoming channel transit to energetically lower outgoing channels, resulting in the loss of ``hot'' atoms from the trap. Additionally, the electric field can induce elastic transitions between different partial waves \cite{prl:96:123202,pra:75:032709,pra:84:032712}. Noting here only the important $s$ and $p$-waves are taken into account. Figure \ref{fig1}(c) depicts four routes involved in inelastic collisions and each of them owns two different pathways. It is essential to note that the incoming channels with two different \( \ell \) values are in the lowest energetic state of \( M_F \), while the outgoing channels are in \( M_F \pm 1 \). In routes 1 and 4, the direct couplings caused by the rf field exhibit imbalanced field-induced couplings, making it challenging to distinguish constructive and destructive patterns. To address this, we propose exploring inelastic scattering via routes 2 and 3, which have balanced field-induced couplings.\\ \indent 
To calculate the two-body loss rate, the time-independent scattering theory considers the complex interaction of magnetic, rf, and electric fields. The corresponding dressed Hamiltonian for the colliding system can be described as follows
\begin{equation}
H=-\frac{\hbar^2}{2\mu}\frac{d^2}{dR^2}+\frac{\hat{\ell}^2}{2\mu R^2}+\hat{V}(R)+H_{\rm int}+H_{\rm fm}+H_{\rm rf}
\end{equation}
where $\mu$ represents the reduced mass of colliding particles and $\hat{\ell}$ the orbital angular momentum of the relative motion between two particles, while $R$ signifies the interatomic separation. The interaction potential, denoted as $\hat{V}(R)$, is contingent upon the total electronic spin $S$, 
\begin{equation}
    \hat{V}(R)=\sum_{SM_S}|SM_S\rangle V_S(R)\langle SM_S|,
\end{equation}
where $V_S(R)$ denotes the adiabatic potential in spin state $S$ whose projection on quantisation $z$ axis defined as $M_S$.   
Furthermore, $H_{\rm int}$ characterizes the internal energy of diatomic molecules, encompassing hyperfine and Zeeman interactions, 
\begin{equation}
    H_{\rm int}=\sum_{j=1,2}\zeta_j \hat{i}_j \cdot \hat{s}_j+(g_s\mu_{\rm B} m_{s_j}+g_n\mu_{\rm B}m_{i_j})B,
\end{equation}
where $j$ represents the index of atom and $\zeta$ is the hyperfine constant; $\hat{i}$ and $\hat{s}$ are the nuclear and electronic spins, with their corresponding projecting quantum numbers $m_i$ and $m_s$ on $z$ axis; $g_s$ and $g_n$ are the electronic and nuclear $g$ factors, respectively; $\mu_{\rm B}$ is the Bohr magneton and $B$ the magnetic field intensity with direction along $z$ axis. $H_{\rm fm}$ delineates the interaction of the colliding complex with electric and rf fields \cite{pra:75:032709,njp:12:083031},
\begin{equation}
    H_{\rm fm}=-E(\vec{e}_E\cdot\vec{e}_d)\sum_{SM_S}|SM_S\rangle d_S(R)\langle SM_S|-\sum_{j=1,2}\vec{\mu}_j\cdot\vec{B}_{\rm rf}
\end{equation}
where $\vec{e}_E$ and $\vec{e}_d$ denote the unit vectors of external electric field $\vec{E}$ and the dipole moment $d_S$ in spin state $S$. $E$ is the electric field intensity. $\vec{\mu}_j$ is the magnetic moment of $j_{\rm th}$ atom and $\vec{B}_{\rm rf}$ is the rf field with intensity $B_{\rm rf}$.
$H_{\rm rf}$ corresponds to the Hamiltonian of the rf field,
\begin{equation}
    H_{\rm rf}=\hbar\omega(\hat{a}\hat{a}^{\dagger}-N_0),
\end{equation}
where $\omega$ is the frequency of rf field. $\hat{a}$ and $\hat{a}^{\dagger}$ are the raising and lowering operators, respectively. $N_0$ is the average photon number of the rf field.\\ \indent
The basis is constructed with $\vert s_am_{s_a}i_am_{i_a}s_bm_{s_b}i_bm_{i_b}\rangle\vert\ell m_{\ell}\rangle\vert N\rangle$, wherein $m_{\ell}$ signifies the projection of $\ell$ on the $z$ axis, while $N$ denotes the dressed photon number of the colliding complex in the field. Under proper boundary conditions in short-range, one can obtain the scattering matrix $\textbf{S}$ in the long-range by solving the time-independent Schr\"{o}dinger equation \cite{jcp:13:445,jcp:85:6425}. In this work the transition rate coefficient between different partial-waves with the same threshold is defined as $\gamma$ which is formulated by 
\begin{equation}
    \gamma=\frac{\pi\hbar}{\mu k}|S_{cc}^{\ell\ell'}|^2,
\end{equation}
where $k$ is the wave vector in the channel $c$. The total two-body loss rate constant $K_2$ is
\begin{equation}
  K_2=\frac{\pi\hbar}{\mu k}\sum_{i\neq e}|S_{ei}|^2,
\end{equation}
where index $i$ ranges over all the open channels other than the incoming channel $e$. \\
\section{Results and Discussion}\label{RD}
We consider $^7$Li-$^{41}$K collisions that are mainly characterized by $s$ and $p$ partial waves at ultralow temperatures $T=1$ nK and use $\sigma^-$ light to drive transitions with $\Delta M_F = \pm 1$ and $\Delta N = \pm 1$. The ground singlet and triplet potentials are adopted from Ref. \cite{pra:79:042716}. The interatomic separation \( R \) are varied from from 2.5 \( a_{\rm 0} \) to \( 1.0 \times 10^4 a_{\rm 0} \), where \( a_{\rm 0} \) represents the Bohr radius. The partial-waves with $\ell\geq$ 2 are excluded in the calculations due to negligible contributions to the rate coefficients at ultralow temperature. The colliding atoms are prepared in the lowest channel of $M_F$ = 2 and we concentrate on the inelastic scattering to the outgoing channel in $M_F$ = 1 by absorbing a $\sigma^-$ photon. Our analysis focuses on the spin-exchange block \( M_F \), which varies from -1 to 4, corresponding to \( N \) values ranging from -3 to 2.  \\ \indent
\begin{figure}[h]
\centering
\resizebox{1.0\textwidth}{!}{%
  \includegraphics{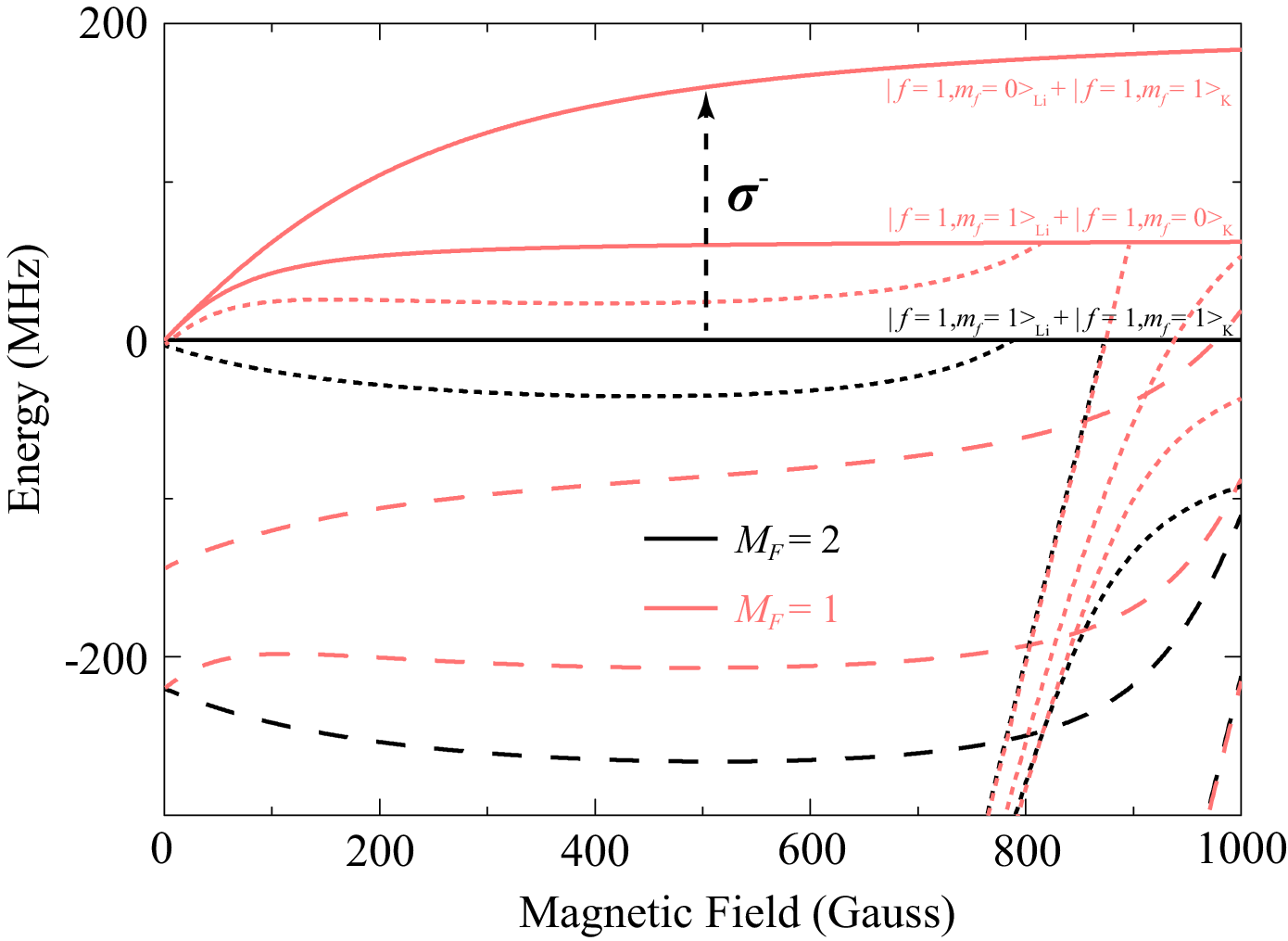}
}\caption{(a) Channel energies (solid lines) and bound states of $^7$Li- $^{41}$K relevant to $M_F$ = 1 and 2 as a function of magnetic field. Long-dashed and dotted lines represent the $s$ and $p$-wave bound states. Notably the bound states correlated to $M_F$ = 1 are displayed below its lowest channel. All the energies are plotted with respect to the colliding channel in $M_F$ = 2. The colliding atoms in the incoming channel can be driven to the outgoing channel by absorbing a $\sigma^-$ photon.}
  \label{fig2}
\end{figure}
 \begin{figure}[h]
\centering
\resizebox{1.0\textwidth}{!}{%
  \includegraphics{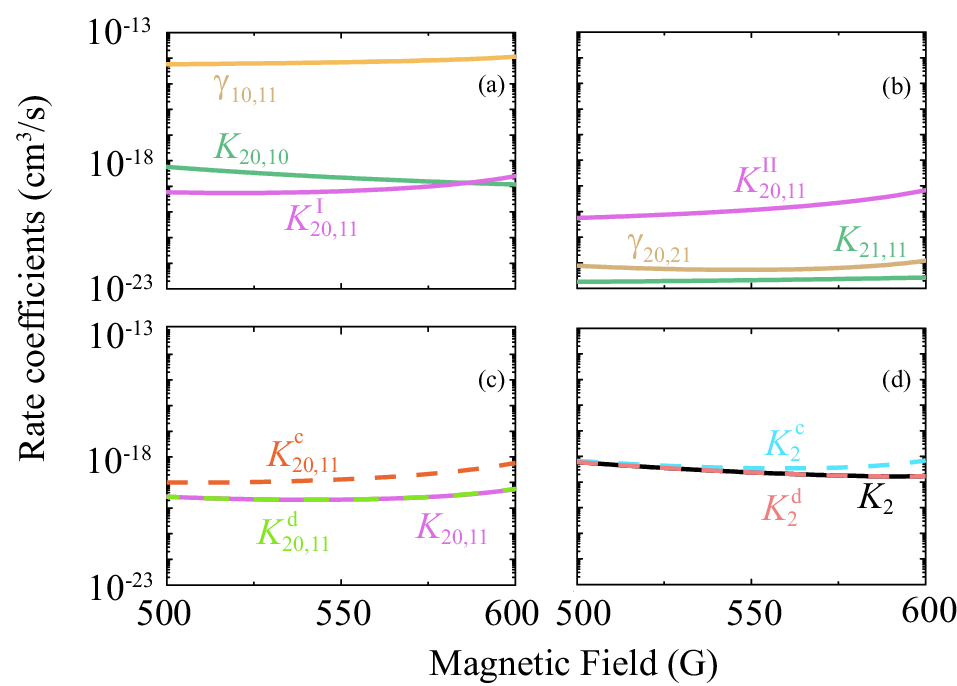}
}\caption{The rate coefficients as a function of magnetic field. The electric and rf field intensities are respective $E$ = 10 kV/cm and $B_{\mathrm{rf}}$ = 0.1 G. The rf frequency is 150 MHz. (a) The partial-wave transition rate coefficient $\gamma_{10,11}$, two-body loss rate coefficients $K_{20,10}$ and $K_{20,11}^{\bf I}$ in pathway $\bf I$. (b) The correlated rate coefficients in pathway $\bf II$. (c) The constructive and destructive rate constants, $K_{20,11}^{\rm c}$ and $K_{20,11}^{\rm d}$, generated with $K_{20,11}^{\bf I}$ and $K_{20,11}^{\bf II}$. $K_{20,11}$ is obtained with coexisting pathways $\bf I$ and $\bf II$. (d) The fully constructive and destructive total two-body loss rates, $K_2^{\rm c}$ and $K_2^{\rm d}$, as well as the practical total two-body loss rate $K_2$.}
  \label{fig3}
\end{figure}
 The rf field plays a key role in our scheme on the visible interference pattern. For optimizing the rf field parameters, we firstly illustrate the correlated channel and bound state energies as a function of the magnetic field in Fig. \ref{fig2}. Due to the outgoing channel not the energetically lowest one in $M_F$ = 1, spin-relaxation process which dominates the inelastic scattering is emerged. In terms of these energies, one can organize the rf-transition paths occurring at the magnetic fields wanted. We state again the inelastic scatterings with balanced field couplings can take place through two possible routes. The first route involves the transition from $s$-wave with $M_F = 2$ to $p$-wave with $M_F = 1$, corresponding to route 2 in Fig. \ref{fig1}(c). The second route involves the transition from $p$-wave with $M_F = 2$ to $s$-wave with $M_F = 1$, corresponding to route 3 in Fig. \ref{fig1}(c), although the latter transition has a low probability due to the $p$-wave scattering at ultracold limit. Our analysis will focus on the first route. This scattering occurs through two intermediate states: $M_F = 2, \ell = 0 \rightarrow M_F = 1, \ell = 0 \rightarrow M_F = 1, \ell = 1$ and $M_F = 2, \ell = 0 \rightarrow M_F = 2, \ell = 1 \rightarrow M_F = 1, \ell = 1$. The following discussion will refer to these pathways as Pathway $\bf I$ and Pathway $\bf II$. \\
\indent 
Figure \ref{fig3} shows the relevant scattering rate coefficients as a function of magnetic field. Here $\gamma_{M_F\ell,M_F\ell'}$ and $K_{M_F\ell,M_F'\ell'}$ describer respective the partial-wave transition and two-body decay rate coefficients. The subscripts $M_F\ell$ or $M_F'\ell'$ express the corresponding incoming or outgoing channel in $\ell$ with $M_F$ or $\ell'$ with $M_F'$. The total two-body loss rate coefficient is represented by $K_2$. The rate coefficients for individual pathways $\bf I$ and $\bf II$ are shown in Figs. \ref{fig3}(a) and (b). At an rf frequency of 150 MHz, the resonances involving either scattering or bound states are far from the magnetic fields presented, resulting in rate coefficients with much lower magnitudes. At ultralow temperatures, quantum effects may cause $K_{20,11}^{\bf I(II)}$ to be larger than the correlated rate coefficients of direct coupling. According to the Landau-Zener theory, $K_{20,11}$ is approximately the product of two field-induced rate coefficients in the classical collision region.\\ \indent
We now examine $K_{20,11}$ in the presence of multiple pathways, where pathways $\bf I$ and $\bf II$ coexist in the calculation. The two-body loss rate coefficient can be expressed in terms of that generated in a single pathway,
\begin{equation}
\begin{aligned}
K_{20,11}=K_{20,11}^{\bf I}+K_{20,11}^{\bf II}+2\cos\theta\sqrt{K_{20,11}^{\bf I}+K_{20,11}^{\bf II}}.
\end{aligned}
\end{equation}
Figure 3(c) plots the fully constructive and destructive loss rate coefficients of $K_{20,11}$, denoted by $K_{20,11}^{\rm c}$ and $K_{20,11}^{\rm d}$, for $\cos\theta$ = 1 and -1, as well as $K_{20,11}$. We can see that  $K_{20,11}$ follows the fully destructive behavior which relies on the scattering phases of incoming and outgoing channels. Figure 3(d) shows the total two-body loss rate $K_2$ and its respective constructive and destructive ones, $K_2^{\rm c}$ and $K_2^{\rm d}$ which are calculated with 
\begin{equation}
\begin{aligned}
K_2^{\rm{c(d)}}=K_{20,11}^{\rm{c(d)}}+K_{20,10}+K_{21,11}+K_{21,10}.
\end{aligned}
\end{equation}
Note we only take into account the influence of $K_{20,10}^{\rm c(d)}$ to $K_2^{\rm c(d)}$ since the tiny differences between constructive and destructive interactions for the left three routes. As $K_{20,10}$ is comparable with $K_{20,11}$, which are two dominated components in total two-body loss rate, $K_2$ exhibit an indiscernible discrepancy between 
two interference profiles. Besides, the such small magnitude order of $K_2$ makes the interference pattern difficult to be verified in experiment.\\ \indent 
\begin{figure}[h]
\centering
\resizebox{0.8\textwidth}{!}{%
  \includegraphics{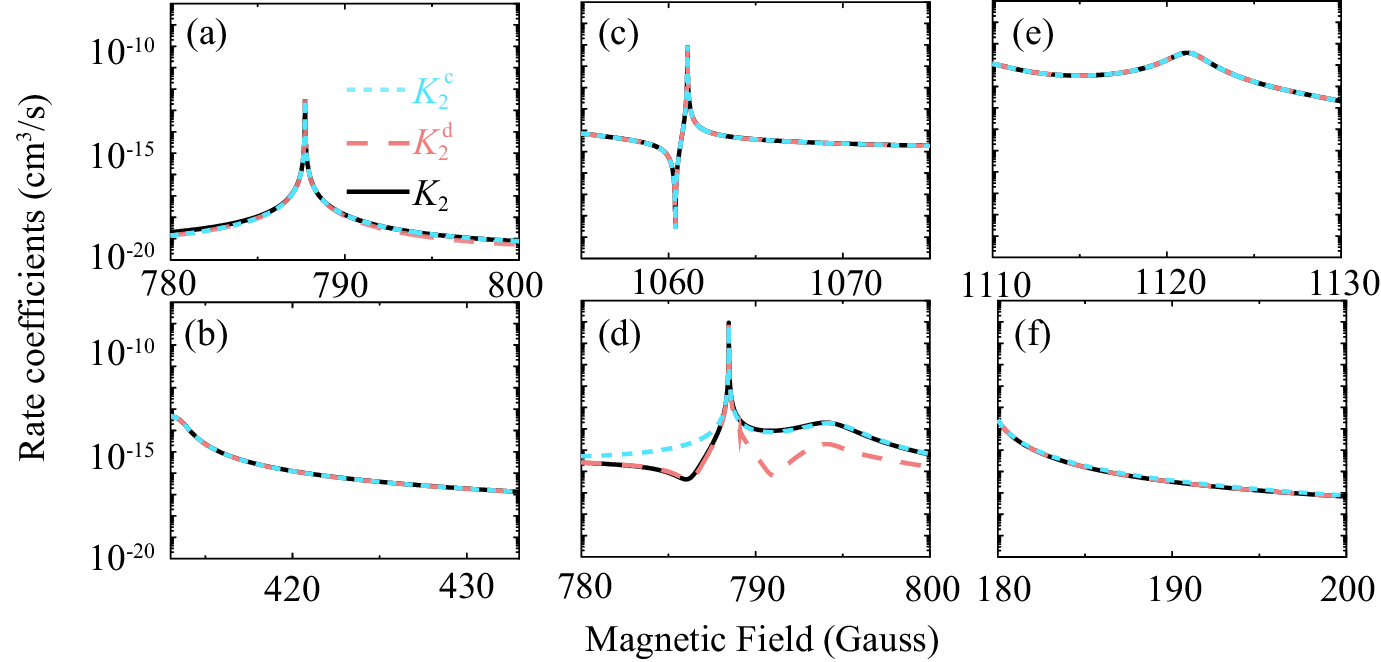}
}\caption{The rate coefficients as a function of magnetic field. $K_2^{\rm c}$ and $K_2^{\rm d}$ represent the fully constructive and destructive total two-body loss rates, respectively. Here the rf frequencies are 150, 150, 185, 176, 185.8 and 97 MHz with rf-induced resonant positions 788, 413, 1050, 763, 1090 and 175 G for corresponding (a)-(f). The electric and rf field intensities are respective $E$ = 10 kV/cm and $B_{\mathrm{rf}}$ = 0.1 G. The magnetically induced resonant positions are 1060.53, 788.31, 1120.94 and 188.78 G corresponding to (c)-(f).}
  \label{fig4}
\end{figure}

\begin{table*}[!t]
    \centering
    \setlength{\tabcolsep}{20mm}{
    \begin{tabular}{@{}ccc@{}}
 \hline\hline
  $|f_{\rm Li},m_{f_{\rm Li}}\rangle$+$|f_{\rm K},m_{f_{\rm K}}\rangle$             & $\ell$ & $B_{\rm 0}$ (G)  \\
  \hline
  $|1,1\rangle$+$|1,1\rangle$     & 0              & 1060.53,\,1076.67 \\
  $|1,1\rangle$+$|1,1\rangle$     & 1              & 788.31,\,872.95   \\
  $|1,0\rangle$+$|1,1\rangle$     & 0              & 1120.94         \\
  $|1,0\rangle$+$|1,1\rangle$     & 1              & 188.78,\,799.58   \\
   \hline\hline
    \end{tabular}}
    \caption{The calculated positions of magnetically induced Feshbach resonances in the absence of electric and rf fields with $B$ $\leq$ 1200 G.}
    \label{tab1}
\end{table*}
In order to address this challenge, we propose leveraging field-induced resonances to augment the rate coefficients \cite{njp:12:083031}. By using particular rf frequencies and magnetic fields, one can realize kinds of rf-induced resonances, including free-to-free, free-to-bound, bound-to-free and bound-to-bound transitions. Whereas magnetically induced Feshbach resonances generally lead to free-to-bound transition. In Tab. \ref{tab1} we list the positions of all the magnetically induced resonances in the relevant channels. Initially, we examined the two-body loss rates near two distinct resonant regions steered by rf field: the free-to-bound and free-to-free transitions. These are illustrated in Fig. \ref{fig4} (a) and (b) where the rf-induced resonances are located around 788 and 413 G, respectively. Both resonant transitions significantly amplify inelastic scattering in vicinity of the resonances; however, the similar values of $K_2^{\rm c}$ and $K_2^{\rm d}$ indicate an interference interaction that is not usable. It is because that the loss from incoming to outgoing channels generally follows the Pathway $\bf I$ leading to $K_{20,11}^{\textbf{I}}$ much larger than $K_{20,11}^{\textbf{II}}$. We then investigated the loss rate close to combined resonant regions, which involve resonances such as rf-induced free-to-free combined with magnetically induced \(s\)-wave with $M_F$ = 2, \(p\)-wave with $M_F$ = 2, \(s\)-wave $M_F$ = 1, and \(p\)-wave with $M_F$ = 1. These transitions can be easily achieved using specific rf frequencies and magnetic fields. Figures \ref{fig4} (c)-(f) display the two-body loss rate coefficients corresponding to these combined resonant transitions. The presence of magnetically induced \(p\)-wave resonance in the incoming channel results in $K_{20,11}^{\textbf{II}}$ comparable with $K_{20,11}^{\textbf{I}}$ near the resonances. As a result, the total two-body loss rate exhibits a clear interference pattern, as shown in Fig. 4(d). The abrupt shift between constructive and destructive interaction arises due to the rapid phase change of the incoming channel's \(p\)-wave scattering wave function. The observable difference between the constructive and destructive rate constants indicates the potential for an interference pattern near the resonance.\\   \indent
\begin{figure}[h]
\centering
\resizebox{0.8\textwidth}{!}{%
  \includegraphics{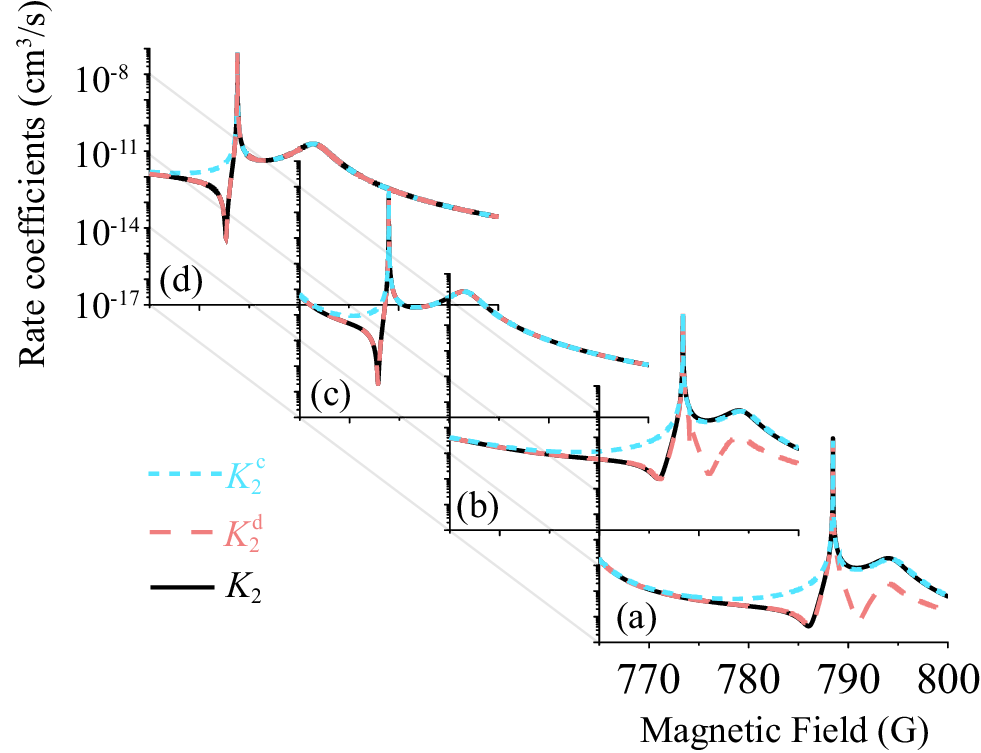}
}\caption{The rate coefficients as a function of magnetic field. $K_2^{\rm c}$ and $K_2^{\rm d}$ represent the theoretically constructive and destructive total two-body loss rates, respectively. Here the rf frequencies are 176 MHz. The external field intensities ($E$, $B_{\rm rf}$) are respective (10 kV/cm, 0.1 G), (10 kV/cm, 0.5 G), (50 kV/cm, 0.1 G) and (50 kV/cm, 0.5 G) corresponding to (a)-(d). }
  \label{fig5}
\end{figure}
The findings above showcase the possibility of detecting inelastic interference by utilizing external rf and electric fields in ultracold collision experiments. However, it should be noted that the total loss rate for most magnetic fields typically falls within the range of \(10^{-15}\) to \(10^{-14}\) \(\rm cm^3/s\) in Fig. \ref{fig4}(d), which may present a challenge for experimentalists. Two strategies can be employed to address this issue without losing the detectability of interference. The first involves increasing the intensities of the external fields, which can enhance both the \(K_{20,10}\) and \(K_{20,11}\) components. Figure \ref{fig5} depicts the two-body loss rates with varying magnetic field strengths for different combinations of rf and electric fields.
In contrast with an electric field, increasing rf intensity leads to detectable interference in a broader magnetic field range and gives more robust applicability. The second strategy is to adjust the rf oscillation frequency to bring the avoided crossing closer to the \(p\)-wave resonance in the incoming channel. This adjustment can intensify the resonance profile, dependent on the background inelastic scattering.\\
\indent 
The contribution from $d$-wave and higher-order partial waves becomes significant when temperatures exceed the ultracold limit. Nevertheless, we focus on rate coefficients at temperatures below 1 $\mu$K, which are typically achievable in current experiments. In this circumstance the interference pattern is not sensitive to temperature. The two-body losses mainly arise from $s$ and $p$-wave interactions, making our findings applicable and reliable. We also conscious that intraspecies collisions may introduce undesired effects. To address this concern, we examined the scattering properties of $^7$Li-$^7$Li and $^{41}$K-$^{41}$K collisions \cite{pra:89:052715,pra:98:022704,pra:98:042708}. Due to the negligible influence of high-order field-complex interactions, we exclude consideration of the electric field in our analysis. Our results show that the rf field can induce a two-body loss rate on the order of $10^{-23}$ cm$^3$/s in $^{41}$K-$^{41}$K collisions within the magnetic fields of interest. On the other hand, in $^7$Li-$^7$Li collisions, the two-body loss is absent due to the energetically lowest incoming channel dressed by rf field. Although elastic scattering may lead to appreciable three-body recombination, its impact on the measurement of the two-body loss rate in interspecies collisions is expected to be negligible. \\ 
\section{Conclusion}\label{Con}
\indent In summary, we showed how external radio-frequency (rf) and electric fields can affect the inelastic scattering process and create a series of ring-coupling structures that enhance our understanding of the interference mechanism. However, verifying interference in the two-body loss rate is challenging when magnetic fields are far from field-induced resonances due to the significant differences in loss rates among the various pathways of inelastic scattering. To address this, we proposed a method to control inelastic scattering using rf and magnetic field-induced resonances. By applying a specific rf field that induces a free-free transition close to the magnetically induced \(p\)-wave Feshbach resonance in the incoming channel, we observed a clear interference pattern in the two-body loss rate around this resonance. Our findings indicate that interference does not necessarily require superposition states. While the rf field-induced avoided crossing between different entrance channels can lead to such states, interference is still observable even in nearly pure states. This suggests that interference is more prominent outside of the avoided crossing. This approach could potentially be applied to other systems that have available magnetically induced \(p\)-wave Feshbach resonances in the incoming channel. \\  
\section{Acknowledgement}
This work was supported by the Innovation Program for Quantum Science and Technology under Grant No. 2021ZD0302101, the National Science Foundation of China under Grants Nos. 22103085 and 12274470, and by the Natural Science Foundation of Hunan Province for Distinguished Young Scholars under Grant No. 2022JJ10070.

\end{document}